\documentclass[twocolumn,floats,showpacs,amsmath,amssymb,pre]{revtex4}
\usepackage{graphicx}
\begin{document}

\title{Kinetic-growth self-avoiding walks on small-world networks}
\author{Carlos P. Herrero}
\affiliation{Instituto de Ciencia de Materiales,
         Consejo Superior de Investigaciones Cient\'{\i}ficas (CSIC),
         Campus de Cantoblanco, 28049 Madrid, Spain }
\date{\today}

\begin{abstract}
Kinetically-grown self-avoiding walks have been studied on Watts-Strogatz 
small-world networks, rewired from a two-dimensional square lattice.
The maximum length $L$ of this kind of walks is limited in regular lattices 
by an attrition effect, which gives finite values for its mean value 
$\langle L \rangle$.  For random networks, this mean attrition length 
$\langle L \rangle$ scales as a power of the network 
size, and diverges in the thermodynamic limit (system size $N \to \infty$).
For small-world networks, we find a behavior that interpolates between 
those corresponding to regular lattices and randon networks, for rewiring 
probability $p$ ranging from 0 to 1.
For $p < 1$, the mean self-intersection and attrition length of
kinetically-grown walks are finite.
For $p = 1$, $\langle L \rangle$ grows with system size as $N^{1/2}$,
diverging in the thermodynamic limit. In this limit and
close to $p = 1$, the mean attrition length diverges as $(1-p)^{-4}$. 
Results of approximate probabilistic calculations agree well with
those derived from numerical simulations.

\end{abstract}

\pacs{89.75.Hc Networks and genealogical trees - 
      05.40.Fb Random walks and Levy flights -
      89.75.Fb Structures and organization in complex systems -
      87.23.Ge Dynamics of social systems} 

\maketitle

\section{Introduction}

In the last few years, researchers have been accumulating evidence that
several kinds of real-life complex systems can be described in terms of
networks or graphs, where nodes represent typical system units and edges
represent interactions between connected pairs of units.
This topological description has been applied to study both natural and man-made 
systems, and is currenlty employed to describe various types of processes taking 
place on real systems (social, biological, economic, technological) 
\cite{st01,al02,do03a,ne03,co05}. 
Two highlights of these developments were the Watts-Strogatz small-world
model \cite{wa98} and the so-called scale-free networks \cite{ba99},
which incorporate several basic ingredients of real systems.

In particular, Watts-Strogatz small-world networks are well suited to study 
properties of systems with underlying topological structure ranging from 
regular lattices to random
graphs \cite{bo98,ca00}, by changing a single parameter \cite{wa99}.
These networks are based on a regular lattice,
in which a fraction $p$ of the links between nearest-neighbor sites
are randomly replaced by new random connections, thus creating long-range 
``shortcuts'' and causing an important decrease in the average distance between 
nodes \cite{al02,ca00}.
In small-world networks one finds at the same time a local neighborhood 
(as for regular
lattices) and some global properties of random graphs \cite{bo98}.
In particular, the ``distance'' between any two elements is small as compared 
with the system size, and the propagation of information 
(or signal, disease, damage, ...) takes place much faster than in 
regular lattices \cite{wa98,al99,la00,ne00a,la01a,la01b,ku01,pa01}.
 This short global length scale affects strongly the behavior
of statistical physical problems on small-world networks,  such as
site and bond percolation \cite{mo00,ne99}, as well as
the Ising model \cite{ba00,gi00,he02a}. 

 Social networks form the substrate where several dynamical processes
take place. These networks have the property of being searchable, 
i.e. people (nodes) can direct messages through their acquaintances to 
reach distant specific targets in only a few steps \cite{wa02,gu02,ki02}.
Although some dynamical processes on complex networks have been studied
by using stochastic dynamics and random walks \cite{je00,ta01,no04},
it is known that real navigation and exploration processes are usually
neither purely random nor totally deterministic \cite{sa01,ad01,ne03}. 
In this context, the generic properties of deterministic navigation \cite{li01} 
and directed self-avoiding walks \cite{sa01} in complex networks have been 
analyzed in the last years.
Self-avoiding walks (SAWs) may be more effective than unrestricted random walks 
in exploring a network, as they cannot return to sites already visited. This 
property has been used to define local search strategies in scale-free 
networks \cite{ad01}.
Nevertheless, the self-avoiding property causes attrition of the paths, since
a large fraction of paths generated in a stochastic way have
to be abandoned because they overlap themselves. This can be an important 
limitation for exploring networks with this kind of walks.

SAWs have been employed for many years for modeling structural and dynamical 
properties of macromolecules \cite{ge79,le89}, as well as 
for studying critical phenomena in lattice models \cite{kr82,pr91}. 
They are also useful to characterize complex crystal structures \cite{he95}
and networks in general \cite{co04}.
In particular, the asymptotic properties of SAWs on small-world networks 
were studied in connection with the so-called {\em connective constant} or
long-distance effective connectivity \cite{he03}.
Recently, kinetic-growth self-avoiding walks on uncorrelated complex networks 
were considered, with particular emphasis upon the influence of attrition on the
maximum length of the paths \cite{he05a,he05b}. It was found that the 
average length scales as a power of the system size $N$, with an exponent that 
depends on the characteristics of the considered networks.
For regular lattices, however, the average maximum
length of kinetic growth SAWs is finite \cite{he84,he86}, contrary to uncorrelated
networks, for which it diverges in the thermodynamic limit ($N \to \infty$). 
Thus, for small-world networks there will appear a crossover when changing
the rewiring probability $p$, from a regime with confined paths
(finite average length typical of regular lattices, $p$ = 0) 
to one with diverging length (characteristic of random networks, $p \to 1$).

Here we study long-range properties of kinetically-grown walks on small-world
 networks, and consider the ``attrition problem''. The number of surviving walks 
to a given length $n$ is obtained by approximate analytical procedures, 
and the results 
are compared with those obtained from numerical simulations. 
We find that the average maximum length increases as the rewiring
probability $p$ is raised, and eventually diverges in the thermodynamic limit
for $p \to 1$, as expected for random networks.
We note that the term {\em length} is employed throughout this paper to
indicate the (dimensionless) number of steps of a walk, as usually done in the
literature on networks \cite{do03a}, and does not correspond to a distance on the
underlying regular lattice.

The paper is organized as follows. 
In Sec.~II we give some definitions and concepts related to SAWs, along
with details on our numerical method.
 In Sec.~III we analyze the length at which non-reversal random walks intersect 
themselves on small-world networks (self-intersection length), and in Sec.~IV 
we calculate the average attrition length of kinetic growth SAWs, at which
they cannot continue without violating the self-avoidance condition.
The paper closes with the conclusions in Sec.~V.
In three appendices we give details of the probabilistic calculations.

\section{Definitions and models}

\subsection{Small-world model}
         
The networks studied here were generated from a two-dimensional square 
lattice (connectivity $z = 4$).
Small-world networks were built up according to the model
of Watts and Strogatz \cite{wa98,wa99}. We consider in turn each
of the bonds in the starting lattice and replace it with a given probability
$p$ by a new bond.  This means that one end of the bond is changed
to a new node chosen at random in the whole network.
We impose the conditions: (i) no two nodes can have more than
one bond connecting them, and (ii) no node can be connected by a link
to itself.
This method keeps constant the total number of links in the
rewired networks, and the average connectivity (or degree) is 
$\langle k \rangle = z$. 
The total number of rewired links is $z p N / 2$ on average.
 In the rewiring process we avoided sites with zero and one links, and thus 
each site has at least two neighbors.
Note that nodes with $k = 1$ behave as culs-de-sac for self-avoiding walks, 
i.e., a walk arriving at a node with connectivity $k=1$ cannot continue,
even though it has not yet intersected itself.

To study the asymptotic behavior of kinetic growth SAWs as $N \to \infty$,
and analyze the convergence of their properties with system size, we
simulated networks of several sizes, including up to $3500 \times 3500$ sites.  
Periodic boundary conditions were assumed.

For small-world networks, the probability distribution of connectivities,
$P_{sw}(k)$, is short-tailed with an exponential decrease for large $k$. 
Analytical expressions for this probability distribution have been
given elsewhere \cite{ba00,he02b}, and will not be repeated here.
We will only give some details necessary for the discussion below.
To this end, we distinguish now between the ends of links that
remain on their original sites (as in the regular lattice),
and those changed in the rewiring process.
The number of links in a network with mean connectivity
$z$ amounts to $z N / 2$. Since each link connecting two
sites in the starting regular lattice is rewired 
with probability $p$, the average number of changes per
site is $z p / 2$. This means that each connection of a
site is removed in the rewiring process with probability $p / 2$.
Then, the number of links $s$ generated in the rewiring 
process and arriving at a node follows, in the limit of large $N$, a
Poisson distribution for $s \geq 0$:
\begin{equation}
P_r(s) = \frac{1}{s!} \, \left( \frac{z p}{2} \right)^s \, e^{-z p/2} \, .
\label{prs}
\end{equation}

We note that other ways of generating small-world networks from regular
lattices have appeared in the literature.
In particular, Newman {\em et al.} \cite{ne99,ne00b} instead of rewiring 
each bond with probability $p$, added shortcuts between
pairs of sites chosen uniformly at random, without removing any bonds
from the starting lattice. This procedure turns out to be more convenient
for analytical calculations, but does not keep constant the mean degree
$\langle k \rangle$, which in this case increases with $p$.

\subsection{Self-avoiding walks}
A self-avoiding walk is defined as a walk along the bonds of
a given network which can never intersect itself. The walk is
restricted to moving to a nearest-neighbor site during each step,
and the self-avoiding condition constrains the walk to occupy only
sites which have not been previously visited in the same walk.
The simplest procedure to obtain SAWs consists in generating ordinary
random walks and stop when they arrive at a node already visited.
In regular lattices, there appears a problem with this sampling algorithm,
caused by the rapid attrition for long walks, because the probability of an
$n$-step walk being self-avoiding decays exponentially for large 
$n$ \cite{so95}.
Due to this important limitation, more complex methods based on Monte 
Carlo sampling have been used to generate SAWs with the correct weight, 
and to obtain ensemble averages of various quantities \cite{so95}. 
In general, for networks including nodes with different degrees (contrary to 
usual regular lattices), sampling by employing simple random walks introduces 
a bias in the weight of different SAWs.

 For regular lattices, the number $u_n$ of SAWs starting
from a site has an asymptotic dependence for large $n$ \cite{pr91,mc76}:
\begin{equation}
         u_n \sim n^{\gamma - 1}   \mu^n  \hspace{2mm} ,
\label{un}
\end{equation}
where $\gamma$ is a critical exponent which depends on the lattice dimension,
and $\mu$ is the {\em connective constant} or effective connectivity
of the considered lattice \cite{mc76,ra85}.
In general, for a lattice with connectivity $z$, one has $\mu \le z - 1$.
 This parameter $\mu$ can be obtained from Eq.~(\ref{un}) as the limit
$\mu = \lim_{n\to\infty} u_n/u_{n-1}$.
The connective constant depends upon the particular topology of each
lattice, and is known with high accuracy for two- and 
three-dimensional lattices \cite{so95,gu01}.
Self-avoiding walks on small-world networks have been studied in recent
years \cite{he03}.
In particular, for networks generated from the square lattice, the
connective constant $\mu$ was found to rise from $\mu = 2.64$ for 
rewiring probability $p = 0$ (regular lattice) to $\mu = 3.70$ for $p=1$.

\begin{figure}
\vspace{-2.8cm}
\includegraphics[width= 9cm]{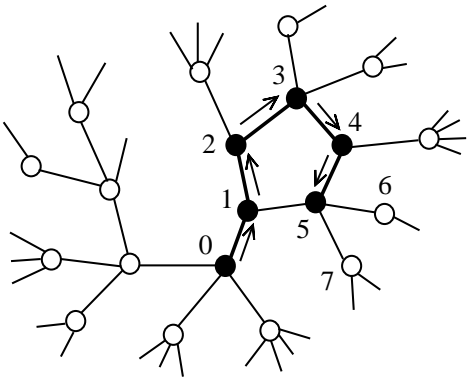}
\vspace{-3.7cm}
\caption{
Schematic diagram showing a non-reversal random walk of length $n = 5$ on a
network. Open and black circles represent unvisited
and visited nodes, respectively. The starting node is indicated as 0.
After arriving at site 5, the non-reversal condition allows to choose for
the next (sixth) step one of three possible nodes (labeled as 1, 6, and 7).
For a non-reversal SAW, one chooses among these three nodes. If 1 is
selected,
then the walk stops.
For a kinetic growth walk, one chooses 6 or 7, each with 50\% probability.
} \label{f1} \end{figure}

To analyze dynamical processes on networks, it can be more convenient to use
kinetically grown SAWs. These walks are well-suited to study
search or navigation processes on networks, where they are assumed to
grow step by step in a temporal sequence.
Their asymptotic behavior can give us also information about the long-range
properties of complex networks, in particular on the presence of loops.
In conection with this, an interesting property of kinetically grown SAWs
is the number of steps necessary to intersect themselves. Moreover, we are
interested in the possibility of avoiding visited nodes and maximize the
length of the walks. 
For these reasons, we will consider in the following two kinds of growing walks. 
The first kind will be {\em non-reversal} self-avoiding walks \cite{so95}. 
In these walks one randomly 
chooses the next step among the neighboring nodes, excluding the previous 
one. If one chooses a node already visited, then the walk 
stops (see Fig. 1). These walks will allow us to study the {\em self-intersection 
length} (see Sec. III).  
The second type of walks considered here are {\em kinetic growth} walks
\cite{ma84}, in which one randomly selects the next step among the
unvisited adjacent sites and stops growing when none are available. 
These walks have been studied earlier to describe the irreversible growth 
of linear polymers \cite{ma84}, and will allow us to consider the {\em attrition
length} for a walk on a given network (see Sec. IV).   
We note that kinetic growth walks are less sensitive to attrition than
non-reversal SAWs, since in the former the walker always escapes whenever 
a way exists. 

For each set of parameters ($p$, $N$), we considered different
network realizations, and for a given network we selected at random the
starting nodes for the SAWs.  
For a given parameter set, the total number of generated SAWs amounted 
to $10^6$ for each type of walks.

\section{Self-intersection length}

To study the probability of a walk intersecting itself, we consider
non-reversal walks that stop when they reach a node already visited 
in the same walk.
The number of steps of a given walk before intersecting itself will be called
{\it self-intersection length} and will be denoted $l$.

Let us consider $M_0$ non-reversal walks starting from nodes taken at
random, and call $M(n)$ the number of walks surviving after $n$ steps
(i.e., those which did not arrive yet at any node visited earlier).
In regular lattices, it is known that the ratio $M(n)/M_0$ decreases
with the number of steps $n$ as $n^{\gamma - 1} e^{-\lambda_0 n}$,
where $\gamma$ is the same exponent as in Eq.~(\ref{un}), and 
$\lambda_0 = \log [(z - 1)/\mu]$.  Hence, the surviving probability 
basically decays for large $n$ as $\exp(-\lambda_0 n)$, with 
$\lambda_0 = 0.129$ for the two-dimensional square lattice \cite{so95}.

\begin{figure}
\vspace{-2.0cm}
\includegraphics[width= 9cm]{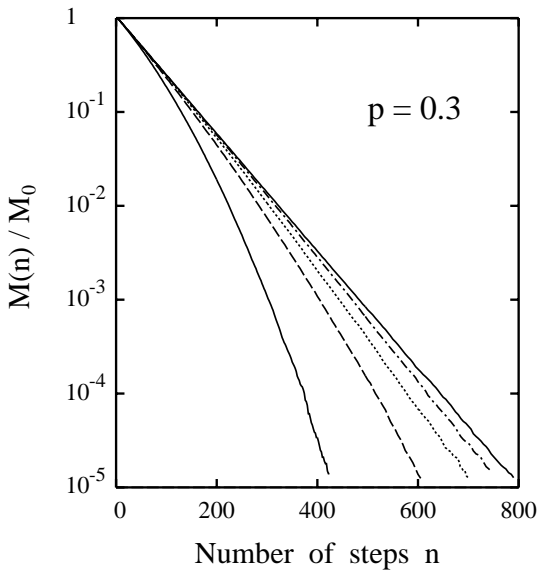}
\vspace{-2.5cm}
\caption{
Fraction of non-reversal SAWs that survive after $n$ steps,
without intersecting themselves. Results are plotted for small-world
networks with rewiring probability $p = 0.3$ and different sizes, as derived
from numerical simulations.
From left to right: $L$ = 100, 200, 300, 500, and 2000.
} \label{f2} \end{figure}

We now turn to the results of our simulations for small-world networks.
In Fig. 2 we show the fraction of remaining walks, $M(n)/M_0$, as a
function of $n$ for networks with several sizes and rewiring probability 
$p = 0.3$.  The network size increases from left to right, and  
one clearly observes a finite-size effect on the surviving probability of 
the walks.
For increasing network size, the results converge to an exponential decrease
of the ratio $M(n)/M_0$. In fact, for the maximum
size presented in Fig. 2 ($2000 \times 2000$ nodes), we find
$M(n)/M_0 = \exp(-\lambda n)$, with $\lambda = 0.014$.
For other $p$ values, one finds similar decays of the number of
surviving walks, with the coefficient $\lambda$ decreasing as $p$ rises.

Let us call $t_n$ the conditional probability of a walk stopping at
step $n+1$, assuming that it actually arrived at step $n$. 
The exponential decay observed for $M(n)/M_0$ is compatible with 
a constant $t_n$, independent of $n$ (i.e., $t_n = t$), as is expected
at least for the asymptotic regime at large $n$.
In fact, assuming a constant $t_n$, one has
\begin{equation}
   M(n) - M(n+1) = t M(n)  \; ,
\label{m1n}
\end{equation}
which can be solved by iteration with the initial condition $M(0) = M_0$,
giving
\begin{equation}
 M(n) = M_0  (1 - t)^n \; .
\label{mm}
\end{equation}
Comparing this result with the exponential decay of $M(n)/M_0$ found above,
we can identify: $\lambda = - \log (1-t)$.

\begin{figure}
\vspace{-2.0cm}
\includegraphics[width= 9cm]{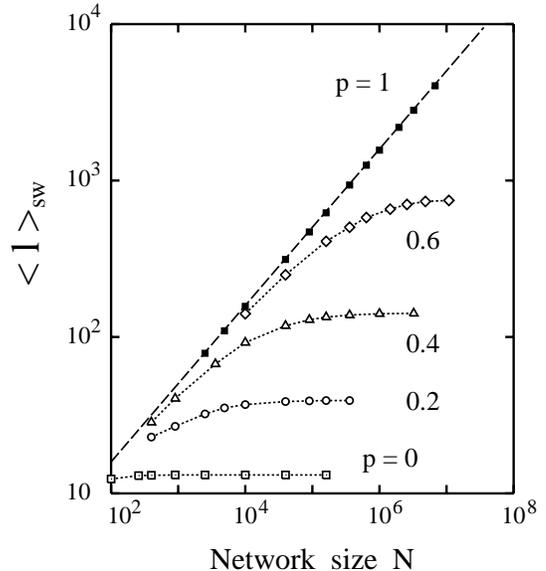}
\vspace{-2.5cm}
\caption{
Average self-intersection length $\langle l \rangle_{sw}$ as a function of
system size $N$ for small-world networks with several rewiring
probabilities $p$.  Symbols display results of numerical simulations:
open squares, $p = 0$ (regular lattice); circles, $p = 0.2$; triangles,
$p = 0.4$; diamonds, $p = 0.6$; filled squares, $p = 1$.
Error bars are less than the symbol size.  Dotted lines are guides to
the eye. The dashed line shows an analytical prediction
for random networks, given by Eq.~(\ref{avl2}).
} \label{f3} \end{figure}

The mean self-intersection length $\langle l \rangle_{sw}$ in small-world
networks is plotted in Fig. 3 
as a function of the system size $N$ for several values of $p$. For a given $N$, 
$\langle l \rangle_{sw}$ increases as $p$ is raised, as could be expected.
For a given rewiring probability $p$, $\langle l \rangle_{sw}$ rises with the
system size, and eventually saturates to a finite value for large $N$.
This saturation appears at larger system sizes as $p$ is
increased. Finally, for $p=1$ we find a behavior similar to that expected
for random networks, for which $\langle l \rangle_{rn}$ increases as a power of
$N$ \cite{he05b}:
\begin{equation}
 \langle l \rangle^2_{rn} \approx \frac{\pi N}{2 w} \; ,
\label{avl2}
\end{equation}
where $w$ is a network-dependent parameter,
$w = (\langle k^2 \rangle - 2 \langle k \rangle) / \langle k \rangle^2$.
The dashed line in Fig. 3 corresponds to Eq.~(\ref{avl2}) for a random network
with minimum degree $k_0 = 2$ and $\langle k \rangle = 4$, which gives 
$\langle l \rangle_{rn} = 1.59 N^{1/2}$  \cite{he05b}.
Thus, the results of our simulations for small-world networks with $p=1$
agree with those expected for random networks, for which 
$\langle l \rangle_{sw}$ diverges with the system size as $N^{1/2}$.

\begin{figure}
\vspace{-2.0cm}
\includegraphics[width= 9cm]{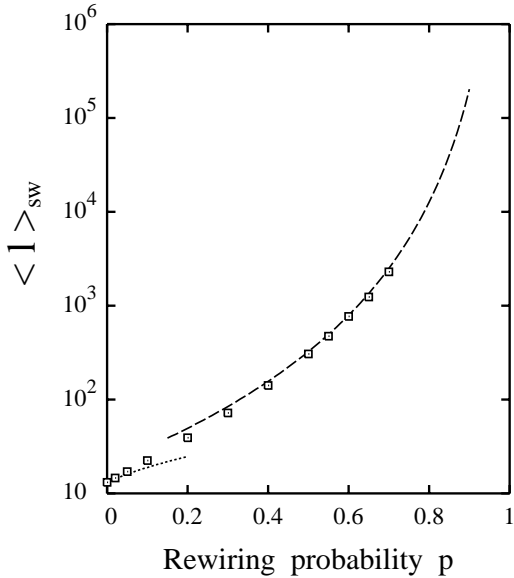}
\vspace{-2.5cm}
\caption{
Mean self-intersection length $\langle l \rangle_{sw}$ as a function of the
rewiring probability $p$ in the thermodynamic limit $N \to \infty$.
 Symbols represent results of numerical simulations. Error bars are less
than the symbol size. Lines indicate results of analytical calculations
for $p \ll 1$ (dotted line) and for $p$ near 1 (dashed line).
} \label{f4} \end{figure}

The mean self-intersection length in the thermodynamic limit is presented
in Fig. 4, as derived from the results of our simulations (open symbols).
In the remainder of this section, we present two analytical approximations
for $\langle l \rangle_{sw}$, valid for the regions close to
$p=0$ and $p=1$. These approximations correspond to the lines shown in
Fig. 4. 

To derive an approximate expression for the mean self-intersection length for
finite (small) $p$, we consider non-reversal SAWs on the regular lattice,
and calculate the corrections introduced by the rewired links in small-world
networks.
An important parameter in this calculation is the conditional probability
$q$ of following a rewired link in step $n+1$, assuming that in fact the walk 
reached step $n$.
To first order in $p$, this probability is given by (see Appendix A):
\begin{equation}
  q =  \frac{p}{2} \left( 1 + \frac{4}{z - 1} \right)  \; ,
\label{qq2}
\end{equation}
where the term $z - 1$ refers to the number of available connections 
for a given step $n$ of a non-reversal walk on the regular lattice.
For the derivation presented in Appendix A, we assume:
(1) The walks include zero or one rewired links, which is compatible
with our order-$p$ approximation, and
(2) a rewired link takes the walker far away from the starting node,
which is valid in average for large system size.

With these assumptions, we obtain for small $p$ Eq.~(\ref{meanl8}), i.e.
\begin{equation}
\langle l \rangle_{sw} \approx \langle l \rangle_{2D} + \frac{q}{2} \left[
       \langle l \rangle_{2D} +  2 \langle l \rangle_{2D}^2 -
       \langle l^2 \rangle_{2D}  \right]    \; ,
\label{meanl6}
\end{equation}
where the subscript $2D$ indicates that the average values are taken for
walks on the square lattice.
Thus, the mean self-intersection length of non-reversal walks on small-world
networks with $p \ll 1$ is expressed in terms of the rewiring probability $p$
(related with $q$ by Eq.~(\ref{qq2})) and known results of the regular lattice.
The mean length $\langle l \rangle_{sw}$ given by Eq.~(\ref{meanl6}) is 
plotted in Fig.~4 as a dotted line.

\begin{figure}
\vspace{-4.7cm}
\includegraphics[width=10cm]{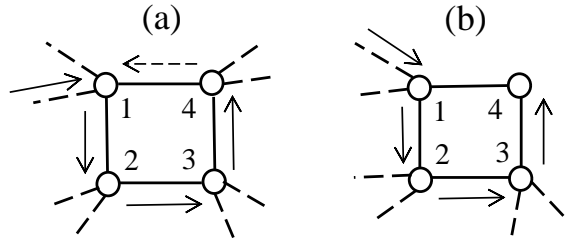}
\vspace{-4.3cm}
\caption{
Schematic representation of SAWs along four-node loops in small-world
networks.
(a) A non-reversal walk arriving at site 1 and following through
sites 2, 3, and 4. The walk stops if it tries to return to site 1 (dashed
arrow).
(b) A kinetic-growth walk arriving at a four-node loop including a site with
$k=2$ (labeled as 4). The walk stops if it follows the sequence of nodes 1,
2, 3, 4, since at node 4 it cannot continue.
} \label{f5} \end{figure}

We now consider the average self-intersection length for rewiring probability
$p$ near 1. As shown above, at $p=1$ we find for small-world networks
the behavior typical of random networks, with
$\langle l \rangle_{rn}$ increasing as $N^{1/2}$, and eventually diverging in the
thermodynamic limit.
For small-world networks with $p < 1$, $\langle l \rangle_{sw}$ has a finite
limit for $N \to \infty$, as shown in Fig.~3, and its value is determined by
correlations present in the networks. Even though these correlations are
small for $p$ close to 1, they still control the long-range behavior of
SAWs. In fact,
in the limit $p \to 1$ and for large enough $N$, the factor limiting the
length of the walks is the residual presence of loops of size four in the 
networks, remaining from the starting regular lattice.
Thus, we assume that the walks finish after circulating along the links of
four-node loops (see Fig. 5(a)), where no one of the connections forming them
was rewired. Taking into account that in the square lattice, the number 
of loops of size 4 equals the number of nodes $N$, then the average number 
$N_S$ of remaining four-node loops is 
\begin{equation}
   N_S = N (1 - p)^4  \; .
\label{ns}
\end{equation}
Note that in the rewiring process new loops of size 4 (and also of size 3) 
can appear, but for large $N$, their number is much less than $N_S$. In fact,
the average number of four-node loops in random networks is basically independent 
of $N$: $N_4 \approx z (z-1)^3 / 2$, and thus for any $p < 1$, $N_S \gg N_4$ for
large enough $N$.
      
With these assumptions, we can calculate the conditional probability $t$ of a
non-reversal SAW stopping at step $n+1$, assuming that it arrived at step
$n$. This is done in Appendix B. 
The probability $t$ turns out to be independent of $n$ for large system size,
and then for $p$ near 1, the probability distribution $Q(l)$ for the 
self-intersection length $l$ can be approximated as
\begin{equation}
   Q(l) = t (1 - t)^l  \; ,
\end{equation}
which gives the probability for a walk proceeding until step $l$, and being
stopped at step $l+1$.

From the distribution $Q(l)$, one finds an average self-intersection length
$\langle l \rangle_{sw} = (1-t)/t$. For $p$ close to 1, one has $t \ll 1$, so
that, using the expressuion for $t$ derived in Appendix B, we find
\begin{equation}
 \langle l \rangle_{sw} \approx \frac{1}{t} = \frac{81}{4} (1 - p)^{-4} \, .
\label{llapr}
\end{equation}
This result takes only into account the self-intersection at loops of size 4,
which gives a dependence of $t$ in the form $(1 - p)^4$. Additional
higher-order terms of the form $(1 - p)^m$ can be obtained by considering
loops of size $m$ in the analyzed networks ($m$ = 6, 8, ...), but their 
contribution will be negligible close to $p = 1$.
The analytical dependence of $\langle l \rangle_{sw}$ given by
Eq.~(\ref{llapr}) is shown in Fig. 4 by a dashed line. It shows good 
agreement with results of the numerical simulations for $p > 0.5$, and predicts
a divergence of $\langle l \rangle_{sw}$ for $p \to 1$, in line with the 
behavior known for random networks. On the other side, for any $p < 1$ our 
calculations predict a finite value for $\langle l \rangle_{sw}$.

It is worth commenting that using the same approximations employed above, 
one finds that the standard deviation of the self-intersection 
length of non-reversal SAWs is $\sigma_l \approx 1/t$, or
$\sigma_l \approx \langle l \rangle_{sw}$. This is consistent with the
exponential behavior of the surviving probability of these walks, as 
discussed above.

\section{Attrition length}

In this section we consider random SAWs, that grow on the network until
they arrive at a node (called {\em blocking node} in the sequel), where they 
cannot continue because all adjacent nodes have been already visited and 
are not available for the walk. These are kinetic growth walks, as 
introduced by Majid {\em et al.}~\cite{ma84}, and defined above in Sec.~II.
The number of steps of a given walk until being blocked will be called
{\em attrition length}, and will be denoted $L$.

\begin{figure}
\vspace{-2.0cm}
\includegraphics[width= 9cm]{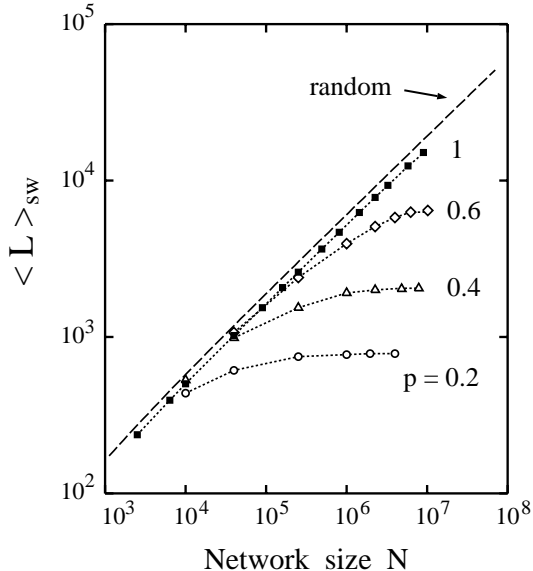}
\vspace{-2.5cm}
\caption{
Mean attrition length $\langle L \rangle_{sw}$ as a function of system size
$N$
for small-world networks with several rewiring probabilities $p$. Symbols
represent results of numerical simulations:
circles, $p = 0.2$; triangles, $p = 0.4$; diamonds, $p = 0.6$;
filled squares, $p = 1$. Error bars are smaller than the symbol size.
Dotted lines are guides to the eye. The dashed line is an approximate
analytical
prediction for random networks, given in Ref.~\cite{he05b} (see text for
details).
} \label{f6} \end{figure}

In Fig. 6 we show the mean attrition length $\langle L \rangle_{sw}$ as a
function of the system size $N$ for several values of the rewiring
probability $p$. Symbols represent results of numerical simulations, 
and dotted lines are guides to the eye.
For a given $N$, $\langle L \rangle_{sw}$ increases as $p$ rises, similarly to
the mean self-intersection length (see Fig. 3). 
For a given rewiring probability $p < 1$, $\langle L \rangle_{sw}$ rises with the
system size, and saturates to a finite value for large $N$.
For $p = 1$, however, we obtain an increase of $\langle L \rangle_{sw}$ as
a power of $N$ in the whole range of network sizes studied here 
(up to $N \approx 10^7$ sites). In fact, we find 
$\langle L \rangle_{sw} \sim N^{\alpha}$, with an exponent 
$\alpha = 0.51 \pm 0.01$.
The dashed line in Fig. 6 represents an analytical approximation for 
the mean attrition length in random networks
(as given by Eqs.~(10) and (24) in Ref.~\cite{he05b}).
According to this calculation, the mean attrition length in uncorrelated 
networks with short-tailed degree distribution scales as 
$\langle L \rangle_{rn} \sim N^{1-1/k_0}$, where $k_0$ is the minimum degree
in the network. For the networks considered here, $k_0 = 2$, and one
should expect $\langle L \rangle_{sw} \sim N^{1/2}$, in line with the results
of our simulations shown in Fig. 6 (filled squares).

We note that the small-world networks with $p=1$ studied here are not
totally uncorrelated, because the restriction $k_0 = 2$ imposes that 
some correlations coming from the starting regular lattice are still present
in the rewired networks. These correlations seem to be unimportant for
defining the main trends of the long-range behavior of kinetic-growth
walks. 

\begin{figure}
\vspace{-2.0cm}
\includegraphics[width= 9cm]{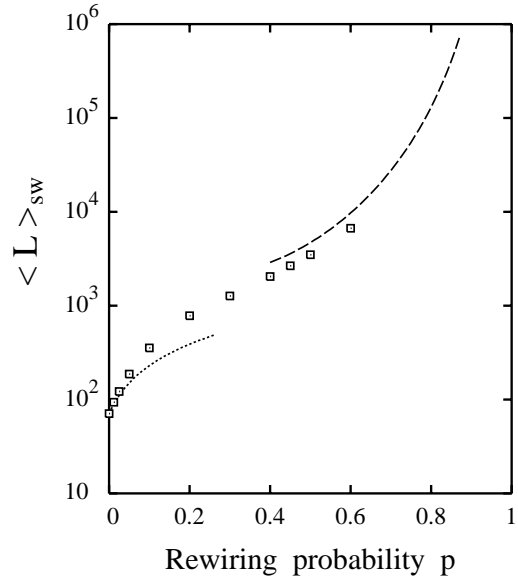}
\vspace{-2.5cm}
\caption{
Mean attrition length $\langle L \rangle_{sw}$ as a function of the
rewiring probability $p$ in the thermodynamic limit $N \to \infty$.
 Symbols represent results of numerical simulations. Error bars are less
than the symbol size. Lines show results of analytical calculations
in the small-$p$ limit (dotted line) and for $p$ close to 1 (dashed line).
} \label{f7} \end{figure}

Shown in Fig. 7 is the mean attrition length in the thermodynamic limit,
as derived from our simulations of kinetic-growth SAWs on rewired networks
(open symbols). For the system sizes considered here, it is hard to obtain
a definite value for the limt $N \to \infty$ for rewiring probabilties $p$
larger than 0.6. Lines in Fig.~7 indicate results of analytical approximations
presented below.

To calculate the mean attrition length of kinetic-growth walks on small-world
networks with $p$ close to cero, we employ a method similar to that used
for the self-intersection length in Sec.~III. In fact, one can follow 
exactly the same steps as in Appendix A, but putting only the connective 
constant $\mu$ instead of $z-1$ in Eqs.~(\ref{p1}) and (\ref{qq}).
In the non-reversal SAWs employed to calculate the
self-intersection length, one can choose among $z-1$ sites at a given step
$n$ on the regular lattice. Instead of this, for kinetic-growth walks used 
to obtain the attrition length, one has in average $\mu$ sites available for
step $n$ (assuming $n \gg 1$). For the square lattice employed here
to obtain the small-world networks, we have $z-1 = 3$ vs $\mu = 2.64$.
Then, we find for the mean attrition length for small $p$:
\begin{equation}
\langle L \rangle_{sw} = \langle L \rangle_{2D} + \frac{q'}{2} \left[
       \langle L \rangle_{2D} +  2 \langle L \rangle_{2D}^2 -
       \langle L^2 \rangle_{2D}  \right]   \; ,
\label{ll1}
\end{equation}
where
\begin{equation}
 q' =  \frac{p}{2} \left( 1 + \frac{4}{\mu} \right)   \; .
\label{qq1}
\end{equation}
The dependence of $\langle L \rangle_{sw}$ on $p$ given by Eqs.~(\ref{ll1}) 
and (\ref{qq1}) is shown in Fig. 7 as a dotted line. As expected, this
approximation to order $p$ gives values of $\langle L \rangle_{sw}$ lower than
the numerical simulations when $p$ is increased from $p=0$. This is due to 
the fact that it neglects higher-order terms, 
with the contribution of walks going on two, three, ..., 
rewired links. This contribution can only increase the
mean attrition length given in Eq.~(\ref{ll1}), which takes into account
walks going over zero or one rewired connection.

To obtain the asymptotic behavior of the mean attrition length 
$\langle L \rangle_{sw}$ close to $p = 1$, we follow a procedure analogous 
to that employed to calculate $\langle l \rangle_{sw}$. Here, as above, 
$\langle L \rangle_{sw}$ is limited by the presence of loops in the 
network, and in particular by loops of size four remaining from the starting 
square lattice.
This means that the highest probability for a kinetic-growth walk being 
blocked occurs at four-node loops, but in this case there appears another 
condition for a path to be stopped, since one needs a blocking node with 
degree $k=2$ (if all four nodes in a loop have $k>2$ the walker will 
escape, see Fig. 5(b)).

With these assumptions, we can calculate the (conditional) probability 
$t'$ of a walk being blocked at step $n+1$, assuming that it actually reached
step $n$. For large $n$, $t'$ can be considered independent of $n$ and
then the distribution of attrition lengths can be expressed as
$Q'(L) = t' (1-t')^L$.   
Finally, the mean attrition length can be approximated for $t' \ll 1$
as $\langle L \rangle_{sw} \approx 1/t'$, and using the expression for 
$t'$ derived in Appendix C, we find
\begin{equation}
 \langle L \rangle_{sw} \approx 27 e^{2p} p^{-2} (1 - p)^{-4} \; .
\label{meanl7}
\end{equation}
This dependence of $\langle L \rangle_{sw}$ on the rewiring probability $p$ 
is displayed in Fig. 7 as a dashed line. The trend predicted by 
Eq.~(\ref{meanl7}) is similar to that of the results of numerical 
simulations in the region $0.4 < p < 0.6$, but somewhat higher (less than a
factor of 2).
Close to $p = 1$ we find a divergence of $\langle L \rangle_{sw}$ as
$(1 - p)^{-4}$, like in the case of the self-intersection length
$\langle l \rangle_{sw}$ (see Eq.~(\ref{llapr})).

For small-world networks rewired from a cubic lattice, 
we expect for $p \to 1$ divergences of the mean self-intersection 
and attrition length similar to those found here
for networks derived from a two-dimensional square lattice.
In fact, for the cubic lattice the minimum-size loops include four
nodes, and those remaining after the rewiring process will control 
$\langle l \rangle_{sw}$ and $\langle L \rangle_{sw}$ close to $p=1$,
which will diverge as $(1 - p)^{-4}$.
Something analogous will happen for networks rewired from hypercubic 
lattices of dimensions higher than three.

\section{Conclusions}
 Self-avoiding walks provide us with an adequate tool to analyze
long-range characteristics of complex networks. In particular,
they allow us to study the quality of a network to be explored without 
returning to sites already visited.
Here, we have studied the self-intersection length $l$ and attrition 
length $L$ of kinetically-grown SAWs on small-world networks, rewired 
from a two-dimensional square lattice.
With this purpose, we have considered non-reversal SAWs to obtain 
$\langle l \rangle_{sw}$ and kinetic-growth walks to obtain 
$\langle L \rangle_{sw}$.

We have calculated self-intersection and attrition lengths by means of
approximate probabilistic methods, which give results in good agreement
with those derived from numerical simulations.
For rewiring probability $p = 1$, both the average self-intersection 
and attrition length diverge with the system size as $N^{1/2}$. 
This dependence of $\langle L \rangle_{sw}$, however, changes with the
minimum degree $k_0$ present in the networks, and in general 
$\langle L \rangle_{sw}$ scales as $N^{\alpha}$ with an exponent 
$\alpha = 1-1/k_0$ (as for random networks).

In the thermodynamic limit, both $\langle l \rangle_{sw}$ and 
$\langle L \rangle_{sw}$ increase with $p$, but remain finite for $p<1$.
The length of kinetically-grown SAWs is limited by the presence of
loops in the networks, mainly those loops remaining from the starting
regular lattice.  Close to $p=1$, both mean lengths diverge as 
$(1-p)^{-4}$. The same type of divergence is expected to appear for
small-world networks rewired from cubic and hypercubic lattices in
higher dimensions.

\begin{acknowledgments}
This work was supported by Ministerio de Educaci\'on y Ciencia (Spain) 
under Contract No. FIS2006-12117-C04-03. \\
\end{acknowledgments}

\appendix
\section{Mean self-intersection length for $p \to 0$}

Here we calculate the mean self-intersection length for small $p$ from
the known results for $p = 0$ (regular lattice). To this end, let us consider
a non-reversal SAW on the regular lattice, that has reached at least
length $n$. Let us call $S_n$ the node visited in step $n$.
For step $n+1$ on the small-world network, we will distinguish three different
possibilities, depending on the nature of the link followed in this step:
                                                                                    
(1) The link employed in step $n+1$ was rewired, but keeping one end on node
$S_n$ (probability $p/2$).

(2) The path follows a link that was rewired from a distant node to node $S_n$.
Since each node is on average connected in the rewiring process to
$p z / 2 = 2 p$ new links,  the conditional probability of
following one of these links in step $n+1$ is
\begin{equation}
  p' =  \frac{2 p}{z - 1 + 2 p}   \; ,
\label{p1}
\end{equation}
which gives to order $p$: $p' \approx 2 p /(z-1)$.
In this expression, $z-1$ is the number of available links for a step of
a non-reversal walk on the regular lattice.
                                                                                    
(3) The link followed in step $n+1$ was not rewired, and remains as in the
starting regular lattice. This occurs with probability $1 - q$, where
$q = p/2 + p'$, or to order $p$:
\begin{equation}
  q =  p \left( \frac12 + \frac{2}{z - 1} \right)  \;  .
\label{qq}
\end{equation}
Then, $q$ is the probability of following a rewired link in step $n+1$,
assuming that in fact the walk reached step $n$.
                                                                                    
Now, let us consider a non-reversal SAW of length $l$ on the regular
lattice, and assume that it intersected itself at step $l+1$.
The same walk will still be possible on a rewired network 
(when no one of the links it uses was rewired) with probability 
$(1 - q)^l$.  For a given $n \leq l$,
the average length of walks going on a rewired connection at step $n+1$
is $n + \langle l \rangle_{2D}$,
where $\langle l \rangle_{2D}$ is the mean self-intersection length in
the regular lattice. This includes the assumption that a
rewired link takes in general the walker to a node on the lattice
far away from the starting one (which is valid in average for large $N$).
We also assume that once a walk has employed a rewired link, it does
not find any other, which is consistent with our order-$p$ approximation for
$\langle l \rangle_{sw}$ (no more than one rewired link in a walk).
Therefore, for walks of length $l$ on the regular lattice, the
average length $\langle l' \rangle_l$ for walks on the rewired networks is:
\begin{equation}
\langle l' \rangle_l = l (1 - q)^l +
                 \sum_{n=1}^l (n + \langle l \rangle_{2D}) q_n    \; ,
\label{meanl2}
\end{equation}
where the second term on the r.h.s. takes into account the contribution of
walks going on a rewired link, and the first, that of walks visiting no
rewired links. Here $q_n$ is the probability of a walk going on a rewired
link at step $n$, given by $q_n = q (1-q)^{n-1}$.

Then, to first order in $q$ we have:
\begin{equation}
\langle l' \rangle_l  \approx  l +  q l \left[ \frac{1-l}{2} +
            \langle l \rangle_{2D} \right]  \; .
\label{meanl5}
\end{equation}
Finally, we calculate the average of $\langle l' \rangle_l$ over all possible
walk lengths $l$ in the regular lattice, and find:
\begin{equation}
\langle l \rangle_{sw} \approx \langle l \rangle_{2D} +  q \left[
       \frac12 \langle l \rangle_{2D} + \langle l \rangle_{2D}^2 -
       \frac12 \langle l^2 \rangle_{2D}  \right]    \;  .
\label{meanl8}
\end{equation}

\section{Mean self-intersection length for $p \to 1$}

Here we present an approximate calculation for $\langle l \rangle_{sw}$ in
small-world networks with $p$ close to 1.
In this limit, and for large enough $N$, the length of the walks is basically 
limited by the residual presence of loops of size four in the networks.
The number of these loops remaining from the starting square lattice is given
by Eq.~(\ref{ns}): $N_S = N (1 - p)^4$.

We then assume that the walks finish after circulating along the links of
four-node loops (see Fig. 5(a)).
Given a loop of size 4 in a small-world network, for each of its nodes
there are in average two
(directed) links leading to the loop from the outside.
Then, the total number of links leading to nodes in these loops is
\begin{equation}
   N_l = 8 N (1 - p)^4  \; .
\end{equation}
  Since the number of directed links  in the network is $4 N$, then the
fraction of those leading to nodes in loops of length 4 is $f = 2 (1 - p)^4$.
Note that we consider nondirected networks, but for our present purpose we
have to distinghish between the two ways in which a single link can be
visited in a walk.

Let us now assume that a walk has arrived at a node (say the node 
labeled ``1'' in Fig.~5(a)) in a four-node loop in step $n-4$ ($n>4$). 
In this case we look for the conditional probability of
circulating around the loop and returning to node ``1'' after four steps,
i.e., in step $n$. This probability $r$ is given by
$r = 2 (1/3)^4 = 2/ 81$, where we have used that the mean connectivity
is $\langle k \rangle = 4$, and that there are two ways to walk around
the considered four-node loop (clockwise and counter-clockwise, see Fig.~5(a)).
Finally, the conditional probability $t$ of arriving in step $n$ at a node
visited earlier (in step $n-4$) is
\begin{equation}
   t = r f = \frac{4}{81}  (1 - p)^4  \; .
\label{tt}
\end{equation}
Note that $t$ defined in Eq.~(\ref{tt}) coincides for $p$ close to 1 with
the stopping probability introduced in Eq.~(\ref{m1n}), and is
independent of the step $n$ (for $n > 4$).

\section{Mean attrition length for $p \to 1$}

  In this appendix we calculate the probability of arriving at a node with 
$k =2$ (labeled as ``4'' in Fig 5(b)) in a four-node loop, after circulating 
along the links of the loop. We assume that node ``4'' has degree two,
so that two of its original links in the regular lattice have
been rewired away (probability $p/2$ for each one), and no other connection
has been established with this node during the rewiring process
(probability $P_r(0)=\exp(-2p)$, see Eq.~(\ref{prs})).
Hence, the average number of nodes with $k=2$ in the remaining four-node
cycles is $N_v = 4 N_S (p/2)^2 P_r(0)$, with $N_S$ given in Eq.~(\ref{ns}).
This yields
\begin{equation}
   N_v = N (1 - p)^4 p^2 e^{-2p} \; .
\label{nv}
\end{equation}

In order for a walk to stop at node ``4'' (in Fig.~5(b)), it has to enter
into the loop through nodes ``1'' or ``3'', and then circulate along the loop
counter-clockwise or clockwise, respectively. Since the average number
of links leading to the loop from the outside through nodes ``1'' or ``3'' 
is four, the fraction of links in the network leading to this kind of
cycles is
\begin{equation}
   f' = \frac{4 N_v}{4 N} = (1 - p)^4 p^2 e^{-2p} \; .
\label{ff1}
\end{equation}
This includes the assumption that the probability of four-node cycles
with more than one node with $k=2$ is negligible (in fact it is much
lower than that of only one node with degree 2).

If a walk has arrived at nodes ``1'' or ``3'' from the outside of the loop,
then the conditional probability of circulating around the loop and
arriving at node ``4'' (the blocking node) is $r' = (1/3)^3$, i.e. $1/3$
for each of three successive steps along the loop.
Thus, similarly to Sec. III and Appendix B, the conditional probability 
$t'$ of arriving 
in step $n$ at a node with $k = 2$, after circulating around a four-node
loop (in the form shown in Fig. 5(b)), is  $t' = r' f'$, or
\begin{equation}
    t' = \frac{1}{27} e^{-2p} p^2 (1 - p)^4 \; .
\label{u1}
\end{equation}

\end{document}